\documentclass[11pt]{article}
\usepackage{amsmath,amssymb,color,epsfig,cite}
\usepackage{graphicx}
\usepackage{setspace}
\usepackage{mathrsfs}
\usepackage[english]{babel}
\usepackage{inputenc}
\usepackage{amsfonts}
\usepackage{soul} 
\usepackage{hyperref}
\usepackage{subfigure}
\usepackage[normalem]{ulem}
\usepackage{array}
\usepackage{authblk}

\newcommand{\hoch}[1]{$\, ^{#1}$}

\makeatletter
\@addtoreset{equation}{section}
\makeatother
\renewcommand{\theequation}{\thesection.\arabic{equation}}

\begin{document}

\vspace{15pt}

{\Large {\bf Homogeneous Projective Coordinates for the Bondi-Metzner-Sachs Group}}
\vspace{15pt}

{\bf Giampiero Esposito \hoch{1,2,}, Giuseppe Filiberto Vitale\hoch{1}}
\vspace{10pt}

\hoch{1}{\it Dipartimento di Fisica ``Ettore Pancini'',
Complesso Universitario di Monte S. Angelo, Via Cintia Edificio 6, 80126 Napoli, Italy}

\hoch{2}{\it INFN Sezione di Napoli, 
Complesso Universitario di Monte S. Angelo, Via Cintia Edificio 6, 80126 Napoli, Italy}

\vspace{20pt}

\underline{ABSTRACT}
This paper studies the Bondi-Metzner-Sachs group in homogeneous
projective coordinates, because it is then possible to write all
transformations of such a group in a manifestly linear way.
The 2-sphere metric, Bondi-Metzner-Sachs metric, asymptotic Killing
vectors, generators of supertranslations, 
as well as boosts and rotations of Minkowski spacetime, are
all re-expressed in homogeneous projective coordinates. 
Last, the integral curves of vector fields which generate supertranslations
are evaluated in detail. This work prepares the ground for more advanced
applications of the geometry of asymptotically flat
spacetimes in projective coordinates, by virtue of the tools
provided from complex analysis in several variables and projective geometry. 

\noindent
\thispagestyle{empty}

\vfill

gesposito@na.infn.it, giuseppef.vitale@gmail.com 

\pagebreak

\section{Introduction}
\setcounter{equation}{0}

The Bondi-Metzner-Sachs \cite{H1,H2,H3} asymptotic symmetry 
group of asymptotically flat spacetime has received again much
attention over the last decade by virtue of its relevance for
black-hole physics \cite{BH1,BH2,BH3}, the group-theoretical
structure of general relativity 
\cite{GT1,GT2,GT3,GT4,GT5,GT6,GT7,GT8,GT9,GT10,GT11,GT12,GT13,GT14}
and the infrared structure of fundamental interactions
\cite{QG0,QG1,QG2,QG3,QG4}.
Moreover, since asymptotic symmetries can provide key constraints
on the celestial dual to quantum gravity in flat spacetimes,
much work has been devoted to the celestial holography
program and related issues \cite{H2,H3,H4,H5,H6,H7,H8}.

The appropriate geometric framework can be summarized as follows. 
In spacetime models for which null infinity can be defined, the 
cuts of null infinity are spacelike two-surfaces orthogonal to the
generators of null infinity \cite{Stewart}. On using the familiar
stereographic coordinate
\begin{equation}
\zeta=e^{i \varphi}\cot {\theta \over 2},
\label{(1.1)}
\end{equation}
the first half of Bondi-Metzner-Sachs transformations read as
\begin{equation}
\zeta'=f(\zeta)={(a \zeta+b)\over (c\zeta+d)}=f_{\Lambda}(\zeta),
\label{(1.2)}
\end{equation}
where the matrix $\Lambda = \left(\begin{matrix}
a & b \cr c & d \end{matrix}\right)$ has unit determinant
$(ad-bc)=1$ and belongs therefore to the group 
${\rm SL}(2,{\mathbb C})$. The resulting projective version
of the special linear group can be defined as the space of pairs
\begin{equation}
{\rm PSL}(2,{\mathbb C})= \left \{ (f,\Lambda)| \;
f: \zeta \in {\mathbb C} \rightarrow
f_{\Lambda}(\zeta),
\Lambda \in {\rm SL}(2,{\mathbb C}) \right \},
\label{(1.3)}
\end{equation}
i.e., the group of fractional linear maps $f_{\Lambda}$ according
to Eq. \eqref{(1.2)} with the associated matrix $\Lambda$. Since
\begin{equation}
f_{\Lambda}(\zeta)={(a\zeta+b)\over (c\zeta+d)}
={(-a \zeta-b)\over (-c \zeta -d)}
=f_{-\Lambda}(\zeta),
\label{(1.4)}
\end{equation}
one can write that ${\rm PSL}(2,{\mathbb C})$ is the quotient
space ${\rm SL}(2,{\mathbb C})/\delta$, where $\delta$ is the
homeomorphism defined by
\begin{equation}
\delta(a,b,c,d)=(-a,-b,-c,-d).
\label{(1.5)}
\end{equation}
The fractional linear maps \eqref{(1.2)} can be defined for all values
of $\zeta$ upon requiring that 
\begin{equation}
f_{\Lambda}(\infty)={a \over c}, \;
f_{\Lambda}\left(-{d \over c}\right)=\infty.
\label{(1.6)}
\end{equation}
Moreover, under fractional linear maps, lengths along the
generators of null infinity scale according to
\begin{equation}
du'=K_{\Lambda}(\zeta)du,
\label{(1.7)}
\end{equation}
where the conformal factor is given by \cite{Stewart,GT13}
\begin{equation}
K_{\Lambda}(\zeta)={{1+ |\zeta|^{2}}\over 
{|a \zeta + b|^{2}+|c \zeta +d|^{2}}}.
\label{(1.8)}
\end{equation}
By integration, Eq. \eqref{(1.7)} yields the second half of
Bondi-Metzner-Sachs transformations:
\begin{equation}
u'=K_{\Lambda}(\zeta)\Bigr[u+\alpha(\zeta,{\overline \zeta})\Bigr].
\label{(1.9)}
\end{equation}
As was pointed out in Ref. \cite{GT13}, the complex homogeneous coordinates
associated to the Bondi-Metzner-Sachs transformation \eqref{(1.2)} 
have modulus $\leq 1$, which is the equation of a unit circle, and are
\begin{equation}
z_{0}=e^{i \varphi \over 2}\cos {\theta \over 2}, \;
z_{1}=e^{-i {\varphi \over 2}} \sin {\theta \over 2}.
\label{(1.10)}
\end{equation}
In other words, upon remarking that
\begin{equation}
\zeta={z_{0}\over z_{1}},
\label{(1.11)}
\end{equation}
Eq. \eqref{(1.2)} is equivalent to the linear transformation law
\begin{equation}
\left(\begin{matrix} z_{0}' \cr z_{1}' 
\end{matrix}\right)
=\left(\begin{matrix}
a & b \cr c & d \end{matrix}\right)
\left(\begin{matrix} z_{0} \cr z_{1} 
\end{matrix}\right).
\label{(1.12)}
\end{equation}
The next step of the program initiated in Ref.
\cite{GT13} consists in realizing that, much in the same
way as the affine transformations in the Euclidean plane
\begin{equation}
x'=x+a, \; y'=y+b,
\label{(1.13)}
\end{equation}
can be re-expressed with the help of a $3 \times 3$ matrix
in the form
\begin{equation}
\left(\begin{matrix}
1 & 0 & a \cr 
0 & 1 & b \cr
0 & 0 & 1
\end{matrix}\right)
\left(\begin{matrix}
x \cr y \cr 1 
\end{matrix}\right)
=\left(\begin{matrix}
x+a \cr y+b \cr 1
\end{matrix}\right),
\label{(1.14)}
\end{equation}
one can further re-express Eq. \eqref{(1.12)} with the help
of a $3 \times 3$ matrix in the form
\begin{equation}
\left(\begin{matrix}
w_{0}' \cr w_{1}' \cr w_{2}'
\end{matrix}\right)
=\left(\begin{matrix}
1 & 0 & 0 \cr
0 & a & b \cr
0 & c & d 
\end{matrix}\right)
\left(\begin{matrix}
w_{0} \cr w_{1} \cr w_{2}
\end{matrix}\right),
\label{(1.15)}
\end{equation}
with the understanding that Eq. \eqref{(1.12)} is the 
restriction to the unit circle $\Gamma$ of the map 
\eqref{(1.15)}, upon defining
\begin{equation}
\left . w_{0} \right |_{\Gamma}=1, \;
\left . w_{1} \right |_{\Gamma}=z_{0}, \;
\left . w_{2} \right |_{\Gamma}=z_{1}. 
\label{(1.16)}
\end{equation}
Within this extended framework, one can consider two complex projective
planes \cite{GT13}. 
Let $P$ be a point of the first plane with coordinates 
$(w_{0},w_{1},w_{2})$, and let $P'$ be a point of the second plane,
with coordinates $(u_{0},u_{1},u_{2})$. One can now consider the nine
products between a complex coordinate of $P$ and a complex coordinate
of $P'$, i.e.
\begin{equation}
Z_{hk}=w_{h}u_{k}, \; h,k=0,1,2.
\label{(1.17)}
\end{equation}
This equation provides the coordinate description of the
{\it Segre manifold} \cite{Caccioppoli,Beltrametti}, 
which is the projective image of the product of projective spaces.
It is a natural tool for accommodating the transformations that 
reduce to the BMS transformations upon restriction to the
unit circle $\Gamma$. It contains a complex double infinity of planes,
two arrays of planes, and a complex fourfold infinity of quadrics
\cite{GT13,Caccioppoli}, but its differential geometry is still
largely unexplored, as far as we know.

Unlike Ref. \cite{GT13}, we have a more concrete
task: since the Bondi-Metzner-Sachs transformation
\eqref{(1.2)} becomes linear when expressed in terms of
$z_{0}$ and $z_{1}$, we are aiming to develop the 
Bondi-Metzner-Sachs formalism with the associated 
Killing vector fields by using the pair
of variables $(z_{0},z_{1})$ instead of
$(\zeta,{\overline \zeta})$. For this purpose, the 
homogeneous projective coordinates for the 2-sphere
are studied in Sect. 2, while the Bondi-Sachs metric
in homogeneous coordinates is considered in Sect. 3. 
Asymptotic Killing fields for supertranslations are 
evaluated in Sect. 4, while their flow is investigated in
Sect. 5. Concluding remarks and open problems are presented
in Sect. 6, while technical details are provided in the
Appendices.

\section{Homogeneous coordinates on the $2$-sphere}

It is useful, as an instrument to develop the BMS formalism in 
homogeneous coordinates, to re-write the 2-sphere metric in the 
desired coordinates. By using the definition (\ref{(1.10)}), we get
\begin{equation}\label{theta}
z_0z_1=\sin(\theta/2)\cos(\theta/2)=\dfrac{\sin(\theta)}{2}
\Rightarrow\theta=\sin^{-1}(2z_0z_1)\text{,}
\end{equation}
while for $\varphi$ we obtain
\begin{equation}
\dfrac{z_0}{z_1}=e^{i\varphi}\cot(\theta/2) 
\Rightarrow \varphi=-i\,\log\left(\tan(\theta/2)\dfrac{z_0}{z_1}\right)\text{.}
\end{equation}
By virtue of the identity
\begin{equation}
\tan(\theta/2) 
=\dfrac{2\sin(\theta/2)\cos(\theta/2)}{2\cos^2(\theta/2)}
=\dfrac{\sin(\theta)}{1+\sqrt{1-\sin^2(\theta)}},
\end{equation}
we obtain for $\varphi$ the more convenient expression
\begin{equation*}
\varphi=-i\,\log\left(\frac{2z^2_0}{1+\sqrt{1-4z^2_0z^2_1}}\right).
\end{equation*}
In order to re-express the $2$-sphere metric, let us evaluate 
\begin{eqnarray}
d\theta^2 &=& \left({\partial \theta \over \partial z_{0}}\right)^{2}
dz_0^2
+\left({\partial \theta \over \partial z_{1}}\right)^{2}
dz_1^2
+2{\partial \theta \over \partial z_{0}}
{\partial \theta \over \partial z_{1}}
dz_0dz_1
\nonumber \\
&=& \frac{4z^2_1}{1-4z^2_0z^2_1}dz_0^2+\frac{4z^2_0}{1-4z^2_0z^2_1}
dz_1^2+\frac{8z_0z_1}{1-4z^2_0z^2_1}dz_0dz_1  \text{,}
\end{eqnarray}
while
\begin{eqnarray}
\sin^2(\theta)\,d\varphi^2 &=& 4z^2_0z^2_1\,d\varphi^2=4z^2_0z^2_1
\left \{ \left({\partial \varphi \over \partial z_{0}}\right)^{2}
dz_0^2+\left({\partial \varphi \over \partial z_{1}}\right)^{2}
dz_1^2
+2{\partial \varphi \over \partial z_{0}}
{\partial \varphi \over \partial z_{1}}
dz_0 dz_1 \right\}
\nonumber \\
&=& \dfrac{-16z^2_1\left(1-2z^2_0z^2_1+\sqrt{1-4z^2_0z^2_1}\right)^2}
{\left(1-4z^2_0z^2_1+\sqrt{1-4z^2_0z^2_1}\right)^2}dz_0^2-
\dfrac{64z^6_0z^4_1}{\left(1-4z^2_0z^2_1+\sqrt{1-4z^2_0z^2_1}\right)^2}dz_1^2
\nonumber \\
&-& \dfrac{32z^3_0z^3_1}{1-4z^2_0z^2_1}dz_0dz_1.
\end{eqnarray}
Eventually, we obtain the metric for the 2-sphere in homogeneous coordinates
\begin{eqnarray}
\Omega_2 &=& d\theta^2+\sin^2(\theta)d\varphi^2=\sum_{\mu,\nu=0}^{1}g_{\mu\nu}dz^{\mu}dz^{\nu}
\nonumber \\
&=& -4z^2_1\left(\frac{1-4z^2_0z^2_1+2\sqrt{1-4z^2_0z^2_1}}{1-4z^2_0z^2_1}\right)
dz_0^2+8z_0z_1dz_0dz_1
\nonumber\\
&-& 4z^2_0\left(\frac{1-4z^2_0z^2_1-2\sqrt{1-4z^2_0z^2_1}}{1-4z^2_0z^2_1}\right)dz_1^2.
\end{eqnarray}
At this stage, upon defining the real-valued function
\begin{equation}
\gamma(z_{0},z_{1})=\frac{2}
{\sqrt{1-4z^2_0z^2_1}}
=\frac{2}{\cos \theta},
\label{(2.7)}
\end{equation}
we can write the matrix of metric components in the form 
\begin{equation}
\gamma_{AB}=
\begin{pmatrix}
-4z^2_1\left(1+\gamma \right) & 4z_0z_1\\                             &                         \\
4z_0z_1 & -4z^2_0\left(1-\gamma \right)
\end{pmatrix},
\end{equation}
with non-vanishing determinant $-16z^2_0z^2_1\gamma^2$ and inverse matrix
\begin{equation}       
\gamma^{AB}=
\begin{pmatrix}
\dfrac{1-\gamma}{4z^2_1\gamma^2} & \dfrac{1}{4z_0z_1\gamma^2}\\                            &                         \\
\dfrac{1}{4z_0z_1\gamma^2} & \dfrac{1+\gamma}{4z^2_0\gamma^2}
\end{pmatrix}.
\end{equation}
We can see from (\ref{theta}) that the terms
\begin{equation*}
2z_0z_1=\sin(\theta)\rightarrow 4z_0^2z^2_1=\sin^2(\theta)\rightarrow 
1-4z^2_0z^2_1=\cos^2(\theta)\rightarrow4z_0z_1=2\sin(\theta),
\end{equation*}
are real-valued, whereas
$$
z^2_0=e^{i\varphi}\cos^2(\theta/2), \; 
z^2_1=e^{-i\varphi}\sin^2(\theta/2)\hspace{1cm}
$$
are complex.

\section{Bondi-Sachs metric in homogeneous coordinates}

We can now write the retarded Bondi-Sachs (hereafter BS) metric 
in homogeneous coordinates with the help of the previous formulae. 
For this purpose, let us first write the general BS metric in the form
\begin{equation}
ds^2=-Udu^2-2e^{2\beta}dudr+h_{AB}\left(dx^A+\frac{1}{2}U^Adu\right)
\left(dx^B+\frac{1}{2}U^Bdu\right).
\label{(3.1)}
\end{equation}
On passing from $(\theta,\varphi)$ to $(z_0,z_1)$ coordinates, we find
the metric components of (3.1) expressed as follows (the material 
from our Eq. (3.2) to our Eq. (3.17) 
can be obtained from Eqs. (4.33), (4.35) and (4.37)
in Ref. \cite{BA1}, which relies in turn upon the work in Ref.
\cite{Tamburino}):
\begin{equation}
g_{uu}=-U+{1 \over 4}h_{z_{0}z_{0}}(U^{z_{0}})^{2}
+{1 \over 4}h_{z_{1}z_{1}}(U^{z_{1}})^{2}
+{1 \over 2}h_{z_{0}z_{1}}U^{z_{0}}U^{z_{1}},
\label{(3.2)}
\end{equation}
\begin{equation}
g_{ur}=-e^{2 \beta},
\label{(3.3)}
\end{equation}
\begin{equation}
g_{uz_{0}}={1 \over 2}(h_{z_{0}z_{0}}U^{z_{0}}
+h_{z_{0}z_{1}}U^{z_{1}}),
\label{(3.4)}
\end{equation}
\begin{equation}
g_{uz_{1}}={1 \over 2}(h_{z_{0}z_{1}}U^{z_{0}}
+h_{z_{1}z_{1}}U^{z_{1}}),
\label{(3.5)}
\end{equation}
\begin{equation}
g_{z_{0}z_{0}}=h_{z_{0}z_{0}}, \;
g_{z_{0}z_{1}}=h_{z_{0}z_{1}}, \;
g_{z_{1}z_{1}}=h_{z_{1}z_{1}}.
\label{(3.6)}
\end{equation}

The Bondi gauge $\partial_r\det\left(r^{-2}g_{AB}\right)=0$ 
implies that 
$\gamma^{AB}C_{AB}=0$, where $\gamma^{AB}$ is given in Eq. (2.9). 
With our coordinates, this relation reads as
\begin{equation*}
\gamma^{AB}C_{AB}=0 \Leftrightarrow g^{z_0z_0}C_{z_0z_0}
+g^{z_1z_1}C_{z_1z_1}+2g^{z_0z_1}C_{z_0z_1}=0.
\end{equation*}
We no longer have the simple result $C_{z\overline{z}}=0$ for the 
mixed component as in the stereographic coordinates, because in 
homogeneous coordinates we obtain
\begin{equation}
\dfrac{1-\gamma}{4z^2_1\gamma^2}C_{z_0z_0}+\dfrac{1+\gamma}
{4z^2_0\gamma^2}C_{z_1z_1}+\dfrac{1}{2z_0z_1\gamma^2}C_{z_0z_1}=0, 
\label{(3.7)}
\end{equation}
which implies that
\begin{equation}
C_{z_0z_1}=-\dfrac{1}{2}\left(1-\gamma\right)\dfrac{z_0}{z_1}C_{z_0z_0}
-\dfrac{1}{2}\left(1+\gamma\right)\dfrac{z_1}{z_0}C_{z_1z_1}.
\label{(3.8)}
\end{equation}

The angular components of the metric are
\begin{equation*}
g_{z_0z_0}=r^2\gamma_{z_0z_0}+rC_{z_0z_0}
+{\mathcal O}(r), \; 
g_{z_1z_1}=r^2\gamma_{z_1z_1}+rC_{z_1z_1}
+{\mathcal O}(r),
\end{equation*}
\begin{eqnarray*}
\; & \; &
g_{z_0z_1}=r^2\gamma_{z_0z_1}+rC_{z_0z_1}
+{\mathcal O}(r)
\nonumber \\
& \; & =r^2\gamma_{z_0z_1}
-r\left(\dfrac{z_0}{z_1}\dfrac{\left(1-\gamma\right)}{2}C_{z_0z_0}
+\dfrac{z_1}{z_0}\dfrac{\left(1+\gamma\right)}{2}C_{z_1z_1}\right)
+{\mathcal O}(r),
\end{eqnarray*}
where, of course, $\gamma_{AB}$ is given in Eq. (2.8). 
These formulae, jointly with the falloff conditions
\begin{align}\label{falloffs}
&U(u,r,x^A)=1-\dfrac{2m(u,r,x^A)}{r}+\dfrac{U_2(u,x^A)}{r^2}+\mathcal{O}(r^{-3})
\nonumber\\
&\beta(u,r,x^A)=\dfrac{\beta_1(u,x^A)}{r}+\dfrac{\beta_2(u,x^A)}{r^2}+\mathcal{O}(r^{-3})
\nonumber\\
&U^A(u,r,x^B)=\dfrac{U^A_2(u,x^B)}{r^2}+\dfrac{U^A_3(u,x^B)}{r^3}+\mathcal{O}(r^{-4})
\nonumber\\
&g_{AB}(u,r,x^A)=r^2\gamma_{AB}(x^A)+rC_{AB}(u,x^A)+D_{AB}(u,x^A)+\mathcal{O}(r^{-1}),
\end{align}
help to rewrite
\begin{equation} \label{guu}
g_{uu}=-\left(1-\dfrac{2m}{r}\right)+\mathcal{O}(r^{-2}).
\end{equation}
Upon assuming that $\beta_1/r\ll1$, we get
\begin{equation}\label{gur}
g_{ur}=-{\rm exp}\left(\dfrac{2\beta_1}{r}+\mathcal{O}
(r^{-2})\right)=-1-\dfrac{2\beta_1}{r}+\mathcal{O}(r^{-2}),
\end{equation}
while for $g_{uz_0}$ and $g_{uz_1}$ we find
\begin{align}\label{guzo}
g_{uz_0}&=\dfrac{1}{2}\left(r^2\gamma_{z_0z_0}+rC_{z_0z_0}\right)
\left(\dfrac{U^{z_0}_2}{r^2}+\dfrac{U^{z_0}_3}{r^3}\right)
+\dfrac{1}{2}\left\{r^2\gamma_{z_0z_1}-r\left[\dfrac{z_0}{z_1}
\dfrac{\left(1-\gamma\right)}{2}C_{z_0z_0}\right.\right.
\nonumber\\
&\left.\left.+\dfrac{z_1}{z_0}\dfrac{\left(1+\gamma\right)}{2}C_{z_1z_1}
\right]\right\}\left(\dfrac{U^{z_1}_2}{r^2}+\dfrac{U^{z_1}_3}{r^3}\right)
\nonumber\\
&=\dfrac{\gamma_{z_0z_0}}{2}U^{z_0}_2+\dfrac{\gamma_{z_0z_1}}{2}U^{z_1}_2
+\dfrac{1}{r}\left[\dfrac{\gamma_{z_0z_0}}{2}U^{z_0}_3
+\dfrac{C_{z_0z_0}}{2}U^{z_0}_2+\dfrac{\gamma_{z_0z_1}}{2}U^{z_1}_3\right.
\nonumber\\
&\left. -\dfrac{z_0}{z_1}\dfrac{\left(1-\gamma\right)}{4}
C_{z_0z_0}U^{z_1}_2-\dfrac{z_1}{z_0}\dfrac{\left(1+\gamma\right)}{4}
C_{z_1z_1}U^{z_1}_2\right]+\mathcal{O}(r^{-2})
\end{align}
and
\begin{align}\label{guzi}
g_{uz_1}&=\dfrac{1}{2}\left(r^2\gamma_{z_1z_1}+rC_{z_1z_1}\right)
\left(\dfrac{U^{z_1}_2}{r^2}+\dfrac{U^{z_1}_3}{r^3}\right)
+\dfrac{1}{2}\left\{r^2\gamma_{z_0z_1}-r\left[\dfrac{z_0}{z_1}
\dfrac{\left(1-\gamma\right)}{2}C_{z_0z_0}\right.\right.
\nonumber\\
&\left.\left.+\dfrac{z_1}{z_0}\dfrac{\left(1+\gamma\right)}{2}
C_{z_1z_1}\right]\right\}\left(\dfrac{U^{z_0}_2}{r^2}
+\dfrac{U^{z_0}_3}{r^3}\right)
\nonumber\\
&=\dfrac{\gamma_{z_1z_1}}{2}U^{z_1}_2+\dfrac{\gamma_{z_0z_1}}{2}U^{z_0}_2
+\dfrac{1}{r}\left[\dfrac{\gamma_{z_1z_1}}{2}U^{z_1}_3
+\dfrac{C_{z_1z_1}}{2}U^{z_1}_2+\dfrac{\gamma_{z_0z_1}}{2}U^{z_0}_3\right.
\nonumber\\
&\left. -\dfrac{z_0}{z_1}\dfrac{\left(1-\gamma\right)}{4}C_{z_0z_0}U^{z_0}_2
-\dfrac{z_1}{z_0}\dfrac{\left(1+\gamma\right)}{4}C_{z_1z_1}U^{z_0}_2\right]
+\mathcal{O}(r^{-2}),
\end{align}
where use has been made of (3.8). Eventually, we get the 
matrix of Bondi metric components
\begin{equation}\label{Bondi}
\hspace{-1cm} g_{\mu\nu}=
\begin{pmatrix}
-\left(1-\dfrac{2m}{r}\right) & -1-\dfrac{2\beta_1}{r} & g_{uz_0} & g_{uz_1}\\\\
-1-\dfrac{2\beta_1}{r}        &     0                  &     0    &      0             \\\\
g_{uz_0}                     &     0                  & r^2\gamma_{z_0z_0}+rC_{z_0z_0}  
& r^2\gamma_{z_0z_1}+rC_{z_0z_1}\\\\
g_{uz_1}          &     0                  &r^2\gamma_{z_0z_1}+rC_{z_0z_1} 
& r^2\gamma_{z_1z_1}+rC_{z_1z_1}
\end{pmatrix}
\hspace{0.2cm}+  \mathcal{O}(r^{-2}).
\end{equation}
The gauge condition ${\rm det}\left(g_{AB}/r^2\right)=0$, instead of 
giving a solution for $D_{AB}$ such as in stereographic coordinates, 
gives us a condition for $C_{AB}$
\begin{equation*}
{\rm det}\left(g_{AB}\right)={\rm det}
\begin{pmatrix}
r^2\gamma_{z_0z_0}+rC_{z_0z_0}+D_{z_0z_0} 
&  r^2\gamma_{z_0z_1}+rC_{z_0z_1}+D_{z_0z_1}\\\\
r^2\gamma_{z_0z_1}+rC_{z_0z_1}+D_{z_0z_1} 
& r^2\gamma_{z_1z_1}+rC_{z_1z_1}+D_{z_1z_1}
\end{pmatrix}
\end{equation*}
\begin{align*}
&=r^4\left(\gamma_{z_0z_0}\gamma_{z_1z_1}-\gamma^2_{z_0z_1}\right)
+r^3\left(\gamma_{z_0z_0}C_{z_1z_1}+\gamma_{z_1z_1}C_{z_0z_0}
-2r^3\gamma_{z_0z_1}C_{z_0z_1}\right)\\
&+r^2\left(\gamma_{z_0z_0}D_{z_1z_1}+C_{z_0z_0}C_{z_1z_1}
+\gamma_{z_1z_1}D_{z_0z_0}-C^2_{z_0z_1}-2\gamma_{z_0z_1}D_{z_0z_1}\right)
+\mathcal{O}(r),
\end{align*}
\begin{align*}
{\rm det}\left(\dfrac{g_{AB}}{r^2}\right)\Rightarrow \,            
&\gamma_{z_0z_0}D_{z_1z_1}+C_{z_0z_0}C_{z_1z_1}
+\gamma_{z_1z_1}D_{z_0z_0}-C^2_{z_0z_1}-2\gamma_{z_0z_1}D_{z_0z_1}\overset{!}{=}0\\
& \Longrightarrow D_{z_0z_1}=\dfrac{\gamma_{z_0z_0}D_{z_1z_1}+C_{z_0z_0}C_{z_1z_1}
+\gamma_{z_1z_1}D_{z_0z_0}-C^2_{z_0z_1}}{2\gamma_{z_0z_1}}\\
&C^2_{z_0z_1}=C_{z_0z_0}C_{z_1z_1}.
\end{align*}
In order to determine the various coefficients in the falloff conditions, we 
require that the Bondi metric should satisfy the Einstein equations
$$
G_{\mu\nu}\equiv R_{\mu\nu}-\dfrac{1}{2}R g_{\mu\nu}=8\pi GT_{\mu\nu}.
$$
Upon restricting to the vacuum case $T=0$, in the limit as $r$ approaches 
$\infty$ in the Einstein tensor, first
looking at $G_{rr}$, and neglecting the terms of order $\mathcal{O}(r^{-4})$, we get 
\begin{equation*}
G_{rr}=-\dfrac{4\beta_1}{r^3}+\mathcal{O}(r^{-4})\overset{!}{=}0\Rightarrow \beta_1\equiv 0.
\end{equation*}
Upon looking at $G_{rz_0}$ and $G_{rz_1}$ respectively, 
we get lengthy relations for $U^{z_1}_2$ and $U^{z_0}_2$, compared to the 
stereographic coordinates case, which depend on other coefficients. 
However, we still manage to solve directly for $U^{z_0}_{2}$ and 
$U^{z_1}_{2}$. On studying $G_{rA}=0$ we find
\begin{equation}
U^{z_0}_2=\dfrac{2z_0z_1\left(C_{z_1z_1}U^{z_1}_2+\gamma_{z_0z_1}U^{z_0}_3
+\gamma_{z_1z_1}U^{z_1}_3\right)}{z^2_1\left(1+\gamma\right)C_{z_1z_1}
+z^2_0\left(1-\gamma\right)C_{z_0z_0}}
=-\dfrac{C_{z_1z_1}U^{z_1}_2+2\gamma_{z_0z_1}U^{z_0}_3}{C_{z_0z_1}},
\end{equation}
and 
\begin{equation}
U^{z_1}_2=\dfrac{2z_0z_1\left(C_{z_0z_0}U^{z_0}_2+\gamma_{z_0z_0}U^{z_0}_3
+\gamma_{z_0z_1}U^{z_1}_3\right)}{z^2_1\left(1+\gamma\right)
C_{z_1z_1}+z^2_0\left(1-\gamma\right)C_{z_0z_0}}
=-\dfrac{C_{z_0z_0}U^{z_0}_2+2\gamma_{z_0z_1}U^{z_1}_3}{C_{z_0z_1}},
\end{equation}
where we recall that $C_{z_0z_1}$ is given in Eq. (3.8). By virtue of
Eqs. (3.12) and (3.13) we find eventually the metric in the form
\begin{eqnarray}
\; & \; & ds^2 =-du^2-2dudr+2\left(r^2\gamma_{z_0z_1}+rC_{z_0z_1}\right)dz_0dz_1
+\dfrac{2m}{r}du^2+\left(r^2\gamma_{z_0z_0}+rC_{z_0z_0}\right)dz^2_0
\nonumber \\
& \; & + \left(r^2\gamma_{z_1z_1}+rC_{z_1z_1}\right)dz^2_1
\nonumber \\
& \; & + \biggr[{\gamma_{z_0 z_0}\over 2}U_{2}^{z_0}
+{\gamma_{z_0 z_1}\over 2}U_{2}^{z_1}
+{1 \over r}\biggr({C_{z_0 z_0}\over 2}U_{2}^{z_0}
+{C_{z_0 z_1}\over 2}U_{2}^{z_1}
+{\gamma_{z_0 z_0}\over 2}U_{3}^{z_0}
+{\gamma_{z_0 z_1}\over 2}U_{3}^{z_1}\biggr)\biggr]du dz_0
\nonumber \\
& \; & + \biggr[
{\gamma_{z_0 z_1}\over 2}U_{2}^{z_0}
+{\gamma_{z_1 z_1}\over 2}U_{2}^{z_1}
+{1 \over r}\biggr(
{C_{z_0 z_1}\over 2}U_{2}^{z_0}
+{C_{z_1 z_1}\over 2}U_{2}^{z_1}
+{\gamma_{z_0 z_1}\over 2}U_{3}^{z_0}
+{\gamma_{z_1 z_1}\over 2}U_{3}^{z_1}\biggr)\biggr]du dz_1
\nonumber \\
& \; & + \mathcal{O}(r^{-2}).
\end{eqnarray}
For the discussion of Bondi's news tensor, mass and angular
momentum aspects we refer again to the work in Refs.
\cite{BA1,Tamburino}.
Now we are ready to evaluate the BMS generators in homogeneous coordinates 
in order to determine the supertranslations.

\section{Asymptotic Killing fields}

After finding the most general Bondi metric in homogeneous coordinates 
satisfying the asymptotically flat spacetime falloffs, our aim is 
to find the most general vector fields $\xi$ satisfying the Bondi 
gauge condition and the asymptotically flat spacetime falloffs.
As is well known, the Killing vectors solve by definition the equations
\begin{equation*}
\left(\mathcal{L}_{\xi}g\right)_{\mu\nu}=\xi^{\rho}\partial_{\rho}
g_{\mu\nu}+g_{\mu\rho}\partial_{\nu}\xi^{\rho}+g_{\nu\rho}\partial_{\mu}\xi^{\rho}=0.
\end{equation*}
Moreover, the preservation of the Bondi gauge condition yields 
\begin{equation}\label{BGc}
\left(\mathcal{L}_{\xi}g\right)_{rr}=0\text{,} \hspace{0.2cm} 
\left(\mathcal{L}_{\xi}g\right)_{rA}=0\hspace{0.2cm} \text{and} 
\hspace{0.2cm} g^{AB}\left(\mathcal{L}_{\xi}g\right)_{AB}=0.
\end{equation}
From these relations one can calculate the four components of $\xi^{\mu}$.
At this stage, we can compute the asymptotic Killing fields in homogeneous  
coordinates by using the familiar transformation law of vector fields.
In other words, the work in Ref. \cite{QG1} has defined the stereographic
variable (we write $\psi$ rather than $z$ used in Ref. \cite{QG1}, 
in order to avoid confusion with our $\zeta$ in Eq. (1.1))
\begin{equation}
\psi=e^{i \varphi}\tan {\theta \over 2}={1 \over {\bar \zeta}}, 
\label{(4.2)}
\end{equation}
and has found, in Bondi coordinates $u,r,\theta,\varphi$, the 
asymptotic Killing fields $\xi_{T}^{+}$ where the components depend
on a function $f$ and on the Bondi coordinates. On denoting as 
usual by $Y_{l}^{m}$ the 
spherical harmonics on the $2$-sphere, one finds 
\begin{equation}
\xi_{T}^{+}\biggr |_{f=Y_{0}^{0}}={\partial \over \partial u},
\label{(4.3)}
\end{equation}
\begin{equation}
\xi_{T}^{+} \biggr |_{f=Y_{1}^{0}}={(1-\psi {\bar \psi})\over (1+\psi {\bar \psi})}
\left({\partial \over \partial u}-{\partial \over \partial r}\right)
+{\psi \over r}{\partial \over \partial \psi}
+{{\bar \psi}\over r}{\partial \over \partial {\bar \psi}},
\label{(4.4)}
\end{equation}
\begin{equation}
\xi_{T}^{+} \biggr |_{f=Y_{1}^{1}}={\psi \over (1+\psi {\bar \psi})}
\left({\partial \over \partial u}-{\partial \over \partial r}\right)
+{\psi^{2}\over 2r}{\partial \over \partial \psi}
-{1 \over 2r}{\partial \over \partial {\bar \psi}},
\label{(4.5)}
\end{equation}
\begin{equation}
\xi_{T}^{+} |_{f=Y_{1}^{-1}}={{\bar \psi}\over (1+\psi {\bar \psi})}
\left({\partial \over \partial u}-{\partial \over \partial r}\right)
-{1 \over 2r}{\partial \over \partial \psi}
+{{\bar \psi}^{2}\over 2r}{\partial \over \partial {\bar \psi}}.
\label{(4.6)}
\end{equation}
Now by virtue of the basic identities
\begin{equation}
{\partial \over \partial \psi}={\partial z_{0}\over \partial \psi}
{\partial \over \partial z_{0}}
+{\partial z_{1}\over \partial \psi}
{\partial \over \partial z_{1}},
\label{(4.7)}
\end{equation}
\begin{equation}
{\partial \over \partial {\bar \psi}}={\partial z_{0}\over \partial {\bar \psi}}
{\partial \over \partial z_{0}}
+{\partial z_{1}\over \partial {\bar \psi}}
{\partial \over \partial z_{1}},
\label{(4.8)}
\end{equation}
and upon exploiting the formulae (A7)-(A10) in the Appendix, we find 
\begin{equation}
\xi_{T}^{+} \biggr |_{f=Y_{1}^{0}}={2 \over \gamma}\left({\partial \over \partial u}
-{\partial \over \partial r}\right)+{z_0 \over 2r}{(2-\gamma)\over \gamma}
{\partial \over \partial z_{0}}
+{z_1 \over 2r}{(2+\gamma)\over \gamma}
{\partial \over \partial z_{1}},
\label{(4.9)}
\end{equation}
\begin{eqnarray}
\xi_{T}^{+} \biggr |_{f=Y_{1}^{1}}&=&
{z_0 \over 2z_1} {(\gamma-2)\over \gamma}
\left({\partial \over \partial u}-{\partial \over \partial r}\right)
+{1 \over r}{(z_0)^{2}\over z_1}\left({1 \over 4}-{1 \over \gamma (\gamma+2)}\right)
{\partial \over \partial z_0}
\nonumber \\
&+& {z_0 \over 2r}\left({1 \over (\gamma+2)}
-{(\gamma+2)\over 2\gamma}\right)
{\partial \over \partial z_{1}},
\label{(4.10)}
\end{eqnarray}
\begin{eqnarray}
\xi_{T}^{+} \biggr |_{f=Y_{1}^{-1}}&=&
{z_1 \over 2z_0}{(\gamma+2)\over \gamma}
\left({\partial \over \partial u}-{\partial \over \partial r}\right)
-{z_1 \over 2r}\left({1 \over (\gamma-2)}
+{(\gamma-2)\over 2 \gamma}\right){\partial \over \partial z_0}
\nonumber \\
&+& {1 \over r}{(z_1)^{2}\over z_0}\left({1 \over 4}
-{1 \over \gamma (\gamma-2)}\right){\partial \over \partial z_{1}}.
\label{(4.11)}
\end{eqnarray}

Now we denote by $\xi_{0},\xi_{1},\xi_{2},\xi_{3}$ the vector
fields (4.3), (4.9), (4.10) and (4.11), respectively. 
Nontrivial Lie brackets among them involve 
$\xi_{1},\xi_{2},\xi_{3}$ only. With our notation, we can
re-write Eqs. (4.9)-(4.11) in the form
\begin{equation}
\xi_{1}=A_{11}\left({\partial \over \partial u}
-{\partial \over \partial r}\right)
+A_{12}{\partial \over \partial z_{0}}
+A_{13}{\partial \over \partial z_{1}},
\label{(4.12)}
\end{equation}
\begin{equation}
\xi_{2}=A_{21}\left({\partial \over \partial u}
-{\partial \over \partial r}\right)
+A_{22}{\partial \over \partial z_{0}}
+A_{23}{\partial \over \partial z_{1}},
\label{(4.13)}
\end{equation}
\begin{equation}
\xi_{3}=A_{31}\left({\partial \over \partial u}
-{\partial \over \partial r}\right)
+A_{32}{\partial \over \partial z_{0}}
+A_{33}{\partial \over \partial z_{1}},
\label{(4.14)}
\end{equation}
where the values taken by the $A_{ij}$ functions can be
read off from (4.9)-(4.11). At this stage, a patient evaluation
proves that such vector fields have vanishing Lie brackets:
\begin{equation}
[\xi_{1},\xi_{2}]=[\xi_{2},\xi_{3}]=[\xi_{3},\xi_{1}]=0.
\label{(4.15)}
\end{equation}
The result is simple, but the actual proof requires several
details, for which we refer the reader to Appendix B.

\begin{figure}[!h]
\centering
\includegraphics[width=11cm]{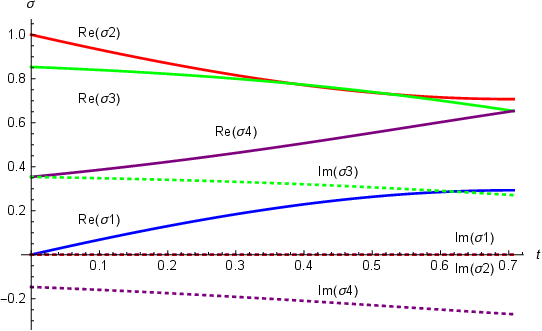}
\caption{Numerical evaluation of the integral curve for the
supertranslation vector field (4.9). 
The initial conditions
(5.14) are taken to be $u=0,r=1,z_{0}=e^{i {\pi \over 8}}
\cos{\pi \over 8},z_{1}=e^{-i{\pi \over 8}}\sin{\pi \over 8}$.}
\end{figure}

\begin{figure}[!h]
\centering
\includegraphics[width=11cm]{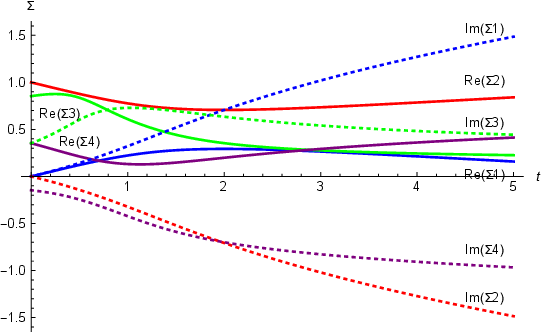}
\caption{Numerical evaluation of the integral curve for the
supertranslation vector field (4.10).
The initial conditions
(5.14) are taken to be $u=0,r=1,z_{0}=e^{i {\pi \over 8}}
\cos{\pi \over 8},z_{1}=e^{-i{\pi \over 8}}\sin{\pi \over 8}$.}
\end{figure}

\begin{figure}[!h]
\centering
\includegraphics[width=11cm]{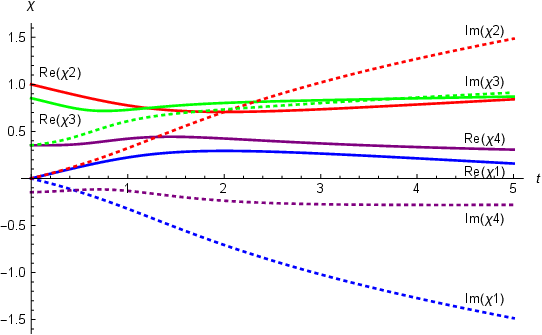}
\caption{Numerical evaluation of the integral curve for the
supertranslation vector field (4.11).
The initial conditions
(5.14) are taken to be $u=0,r=1,z_{0}=e^{i {\pi \over 8}}
\cos{\pi \over 8},z_{1}=e^{-i{\pi \over 8}}\sin{\pi \over 8}$.}
\end{figure}

\begin{figure}[!h]
\centering
\includegraphics[width=11cm]{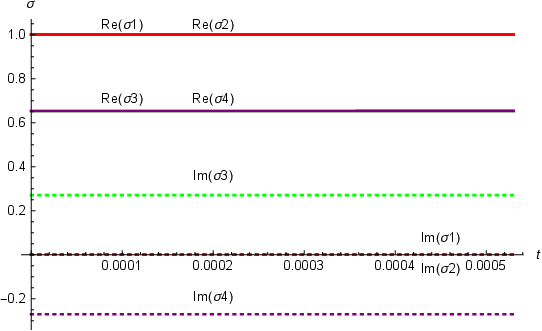}
\caption{Numerical evaluation of the integral curve for the
supertranslation vector field (4.9). 
The initial conditions
(5.14) are taken to be $u=1,r=1,z_{0}=e^{i {\pi \over 8}}
\cos{\pi \over 4},z_{1}=e^{-i{\pi \over 8}}\sin{\pi \over 4}$.}
\end{figure}

\begin{figure}[!h]
\centering
\includegraphics[width=11cm]{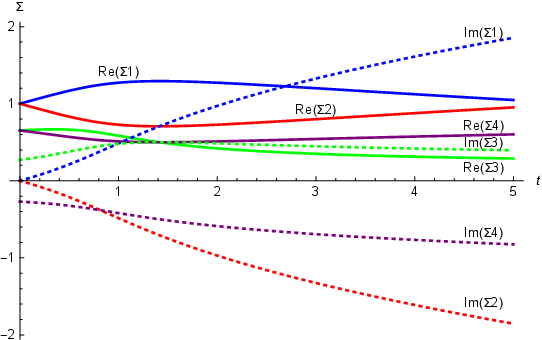}
\caption{Numerical evaluation of the integral curve for the
supertranslation vector field (4.10).
The initial conditions
(5.14) are taken to be $u=1,r=1,z_{0}=e^{i {\pi \over 8}}
\cos{\pi \over 4},z_{1}=e^{-i{\pi \over 8}}\sin{\pi \over 4}$.}
\end{figure}

\begin{figure}[!h]
\centering
\includegraphics[width=11cm]{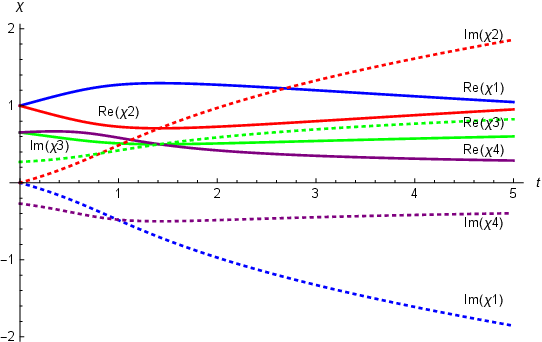}
\caption{Numerical evaluation of the integral curve for the
supertranslation vector field (4.11).
The initial conditions
(5.14) are taken to be $u=1,r=1,z_{0}=e^{i {\pi \over 8}}
\cos{\pi \over 4},z_{1}=e^{-i{\pi \over 8}}\sin{\pi \over 4}$.}
\end{figure}

\begin{figure}[!h]
\centering
\includegraphics[width=11cm]{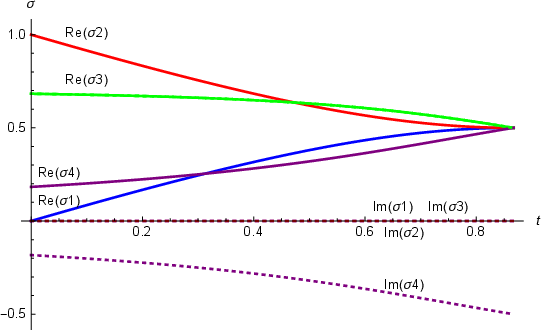}
\caption{Numerical evaluation of the integral curve for the
supertranslation vector field (4.9). 
The initial conditions
(5.14) are taken to be $u=0,r=1,z_{0}=e^{i {\pi \over 4}}
\cos{\pi \over 12},z_{1}=e^{-i{\pi \over 4}}\sin{\pi \over 12}$.
In this particular case, the real parts meet at a single point.}
\end{figure}

\begin{figure}[!h]
\centering
\includegraphics[width=11cm]{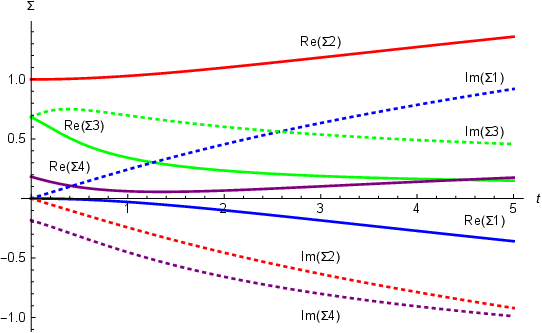}
\caption{Numerical evaluation of the integral curve for the
supertranslation vector field (4.10). 
The initial conditions
(5.14) are taken to be $u=0,r=1,z_{0}=e^{i {\pi \over 4}}
\cos{\pi \over 12},z_{1}=e^{-i{\pi \over 4}}\sin{\pi \over 12}$.}
\end{figure}

\begin{figure}[!h]
\centering
\includegraphics[width=11cm]{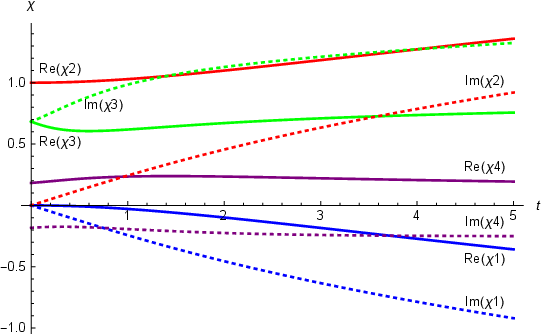}
\caption{Numerical evaluation of the integral curve for the
supertranslation vector field (4.11). 
The initial conditions
(5.14) are taken to be $u=0,r=1,z_{0}=e^{i {\pi \over 4}}
\cos{\pi \over 12},z_{1}=e^{-i{\pi \over 4}}\sin{\pi \over 12}$.}
\end{figure}

\section{Flow of supertranslation vector fields}

The analysis in this section does not have a direct impact
on unsolved problems, but (as far as we can see) can help
the general reader. More precisely,
in order to appreciate that the familiar geometric constructions
are feasible also in projective coordinates, we now consider the
flow of supertranslation vector fields (4.9)-(4.11). For example,
by virtue of (2.7), and defining $p=(u,r,z_{0},z_{1})$, 
the task of finding
the flow of the supertranslation vector fields (4.9), (4.10)
and (4.11) consists of solving a system of nonlinear and coupled  
differential equations. For this purpose, we denote by 
$\sigma,\Sigma,\chi$, respectively, the appropriate flow, and define
\begin{equation}
\delta(W;\tau,p)=\sqrt{1-4 \Bigr(W^{3}(\tau,p)W^{4}(\tau,p)\Bigr)^{2}},
\label{(5.1)}
\end{equation}
where $W=\sigma,\Sigma,\chi$, respectively, with components
$W^{1},W^{2},W^{3},W^{4}$. 
Hence we study the following coupled systems of nonlinear
differential equations:
\begin{equation}
{d \sigma^{1}\over d\tau}=\delta(\sigma;\tau,p),
\label{(5.2)}
\end{equation}
\begin{equation}
{d \sigma^{2}\over d\tau}= - \delta(\sigma;\tau,p),
\label{(5.3)}
\end{equation}
\begin{equation}
{d\sigma^{3}\over d\tau}
={\sigma^{3}(\tau,p)\over 2 \sigma^{2}(\tau,p)}
(\delta(\sigma;\tau,p)-1),
\label{(5.4)}
\end{equation}
\begin{equation}
{d\sigma^{4}\over d\tau}
={\sigma^{4}(\tau,p)\over 2 \sigma^{2}(\tau,p)}
(\delta(\sigma;\tau,p)+1),
\label{(5.5)}
\end{equation}
\begin{equation}
{d \Sigma^{1}\over d\tau}={\Sigma^{3}(\tau,p)\over 2 \Sigma^{4}(\tau,p)}
(1-\delta(\Sigma;\tau,p)),
\label{(5.6)}
\end{equation}
\begin{equation}
{d \Sigma^{2}\over d\tau}=-{\Sigma^{3}(\tau,p)\over 2 \Sigma^{4}(\tau,p)}
(1-\delta(\Sigma;\tau,p)),
\label{(5.7)}
\end{equation}
\begin{equation}
{d \Sigma^{3}\over d\tau}={(\Sigma^{3}(\tau,p))^{2}\over
4 \Sigma^{2}(\tau,p)\Sigma^{4}(\tau,p)}
\left[1-\delta(\Sigma;\tau,p)+{\delta(\Sigma;\tau,p)\over
(1+\delta(\Sigma;\tau,p))}\right],
\label{(5.8)}
\end{equation}
\begin{equation}
{d \Sigma^{4}\over d\tau}={\Sigma^{3}(\tau,p)\over 4 \Sigma^{2}(\tau,p)}
\left[{\delta(\Sigma;\tau,p)\over (1+\delta(\Sigma;\tau,p))}
-1-\delta(\Sigma;\tau,p)\right],
\label{(5.9)}
\end{equation}
\begin{equation}
{d \chi^{1}\over d\tau}={\chi^{4}(\tau,p)\over 2 \chi^{3}(\tau,p)}
(1+\delta(\chi;\tau,p)),
\label{(5.10)}
\end{equation}
\begin{equation}
{d \chi^{2}\over d\tau}=-{\chi^{4}(\tau,p)\over 2 \chi^{3}(\tau,p)}
(1+\delta(\chi;\tau,p)),
\label{(5.11)}
\end{equation}
\begin{equation}
{d \chi^{3}\over d\tau}=-{\chi^{4}(\tau,p)\over 4 \chi^{2}(\tau,p)}
\left[{\delta(\chi;\tau,p)\over (1-\delta(\chi;\tau,p)}
+1-\delta(\chi;\tau,p)\right],
\label{(5.12)}
\end{equation}
\begin{equation}
{d \chi^{4}\over d\tau}={(\chi^{4}(\tau,p))^{2}\over 
4 \chi^{2}(\tau,p) \chi^{3}(\tau,p)} 
\left[1+\delta(\chi;\tau,p)-{\delta(\chi;\tau,p)\over
(1-\delta(\chi;\tau,p))}\right],
\label{(5.13)}
\end{equation}
with the initial conditions
\begin{equation}
W^{1}(0,p)=u, \;
W^{2}(0,p)=r, \;
W^{3}(0,p)=z_{0}, \;
W^{4}(0,p)=z_{1}.
\label{(5.14)}
\end{equation}
The resulting equations can only be solved numerically,
to the best of our knowledge, and such solutions are displayed
in Figures from 1 to 9. Since the
desired solutions are complex-valued, we have displayed
both real and imaginary parts, with three choices of 
initial conditions.

\section{Concluding remarks and open problems}

As far as we can see, the interest of our investigation lies
in having shown that homogeneous projective coordinates lead 
to a fully computational scheme for all applications of the
BMS group. This might pay off when more advanced properties
will be studied. In particular, we have in mind the concept
of superrotations \cite{QG1,QG2} 
on the one hand, and the physical applications
of the Segre manifold advocated in our Introduction and  
in Ref. \cite{GT13}. In other
words, since our Eq. (1.15) contains Eq. (1.12), which in
turn is just a re-expression of the BMS transformation (1.2),
one might aim at embedding the study of BMS symmetries into
the richer mathematical framework of complex analysis 
in several variables \cite{2006} and algebraic geometry.
The exploitment of the complex analysis approach to algebraic
geometry appears promising because the singular points of 
functions of several complex variables form a continuum
(see definitions and theorems in Refs. \cite{Caccioppoli,2006}).
The potentialities of this framework for studying e.g. 
superrotations were unforeseen so far, and deserve
careful consideration in our opinion. 
Our paper has tried to prepare the ground
for such a synthesis, even though our calculations are 
not cumbersome.

Moreover, we would like to mention that the research in Refs.
\cite{BA1,BA2,Freidel} has exploited the fact that one can
actually work with a completely arbitrary metric on the
asymptotic 2-sphere. By doing so, one can write the on-shell
expression of $U,U^{A}$ and $\beta$ in our Sect. 3 in terms
of this arbitrary 2-sphere metric. This might therefore
provide a way to recover our results when taking the particular
case in which the 2-sphere metric is expressed in homogeneous
projective coordinates. We are grateful to M. Geiller for
this remark, and also for having brought to our attention
the work in Ref. \cite{Geiller}, where the authors have
written the solution space and the asymptotic Killing vectors
and their action in the case of an even more general gauge
than Bondi-Sachs.

\section*{Author Contributions} 

The authors have equally contributed to conceptual and
technical parts of the paper. All authors have read
and agreed to the published version of the manuscript.

\section*{Funding}

This research received no external funding.

\section*{Data Availability Statement}

No new data were created.

\section*{Acknowledgments}

The authors are grateful to Professor Patrizia Vitale for
encouraging their collaboration. G. Esposito is grateful
to INDAM for membership.

\section*{Conflicts of Interest}

The authors declare no conflict of interest.

\vskip 100cm

\renewcommand{\theequation}{A.\arabic{equation}}

\leftline {\bf Appendix A: the use of homogeneous coordinates}
\setcounter{equation}{0}

\vskip 1cm

By virtue of Eqs. (1.10) and (2.7), we find
\begin{equation}
(z_0)^{2}=e^{i \varphi}{(1+\cos \theta)\over 2}
\Longrightarrow e^{i \varphi}
={2(z_0)^{2}\over (1+\cos \theta)}
={2(z_0)^{2}\gamma \over (\gamma+2)},
\label{(A1)}
\end{equation}
and hence the variable $\psi$ in Eq. (4.2) can be re-expressed in the form
\begin{equation}
\psi={2(z_0)^{2}\gamma \over (\gamma+2)}
{(1-\cos \theta)\over \sin \theta}
={2(z_0)^{2}\gamma \over (\gamma+2)}
{\left(1-{2 \over \gamma}\right)\over 2z_{0} z_{1}}
={z_0 \over z_1}{(\gamma-2)\over (\gamma+2)},
\label{(A2)}
\end{equation}
while
\begin{equation}
{\bar \psi}={1 \over \zeta}={z_1 \over z_0}.
\label{(A3)}
\end{equation}
Moreover, we need the identities
\begin{equation}
\psi {\bar \psi}={\left(1-\cos^{2} {\theta \over 2}\right)\over
\cos^{2} {\theta \over 2}} \Longrightarrow
\cos {\theta \over 2}
={1 \over \sqrt{(1+\psi {\bar \psi})}}, \;
\sin {\theta \over 2}=\sqrt{\psi {\bar \psi} \over
(1+ \psi {\bar \psi})},
\label{(A4)}
\end{equation}
\begin{equation}
e^{i {\varphi \over 2}}=\left({\psi \over {\bar \psi}}\right)^{1 \over 4}, \;
e^{-i {\varphi \over 2}}=\left({{\bar \psi}\over \psi}\right)^{1 \over 4},
\label{(A5)}
\end{equation}
which, jointly with the definitions (1.10), lead to
\begin{equation}
z_0=\left({\psi \over {\bar \psi}}\right)^{1 \over 4}
{1 \over \sqrt{1+\psi {\bar \psi}}}, \;
z_1=\left({{\bar \psi}\over \psi}\right)^{1 \over 4}
\sqrt{\psi {\bar \psi} \over (1+\psi {\bar \psi})}.
\label{(A6)}
\end{equation}
At this stage, we can evaluate the partial derivatives
occurring in Eqs. (4.7) and (4.8) 
by patient application of Eqs. (A2), (A3) and (A6), i.e.,
\begin{equation}
{\partial z_0 \over \partial \psi}
={z_{1}\over 2}{(\gamma+2)\over \gamma(\gamma-2)},
\label{(A7)}
\end{equation}
\begin{equation}
{\partial z_{1}\over \partial \psi}
={1 \over 2}{(z_1)^{2}\over z_0}
{(\gamma+2)\over \gamma(\gamma-2)},
\label{(A8)}
\end{equation}
\begin{equation}
{\partial z_0 \over \partial {\bar \psi}}
=-{(z_0)^{2}\over 2z_1}{(\gamma-1)\over \gamma},
\label{(A9)}
\end{equation}
\begin{equation}
{\partial z_{1}\over \partial {\bar \psi}}
={z_0 \over 2}{(\gamma+1)\over \gamma},
\label{(A10)}
\end{equation}
and we find eventually the asymptotic Killing fields in the 
form (4.9)-(4.11).
Our homogeneous projective coordinates $z_0$ and $z_1$ have also
been considered in Ref. \cite{Penrose}, but in that case, upon writing
\begin{equation}
\zeta={(x+iy)\over (1-z)},
\label{(A11)}
\end{equation}
one finds that the $x,y,z$ coordinates for the embedding of the
$2$-sphere in three-dimensional Euclidean space are given by
\begin{equation}
x={2 {\rm Re}(\zeta)\over (1+|\zeta|^{2})}
={(z_0 {\bar z}_1 + {\bar z}_0 z_1) \over
(|z_0|^2 + |z_1|^2)},
\label{(A12)}
\end{equation}
\begin{equation}
y={2 {\rm Im}(\zeta)\over (1+|\zeta|^2)}
={(z_0 {\bar z}_{1}-{\bar z}_{0}z_1) \over
i (|z_0|^{2}+|z_1|^{2})},
\label{(A13)}
\end{equation}
\begin{equation}
z={(|\zeta|^{2}-1) \over (|\zeta|^{2}+1)}
={(|z_0|^{2}-|z_1|^2)\over (|z_0|^2+|z_1|^2)}.
\label{(A14)}
\end{equation}

The global spacetime translations of Minkowski spacetime can be
first re-expressed in $u,r,\xi,{\bar \xi}$ coordinates, and
read eventually, in terms of the asymptotic Killing fields (4.9)-(4.11),
\begin{eqnarray}
\; & \; &
X_0=-\xi_{T}^{+} \biggr |_{f=Y_{0}^{0}}, \;
X_{1}=-\xi_{T}^{+} \biggr |_{f=Y_{1}^{1}}-\xi_{T}^{+}\biggr |_{f=Y_{1}^{-1}}, 
\nonumber \\
& \; & 
i X_{2}=\xi_{T}^{+} \biggr|_{f=Y_{1}^{-1}}
-\xi_{T}^{+} \biggr|_{f=Y_{1}^{1}}, \;
X_{3}=-\xi_{T}^{+} \biggr |_{f=Y_{1}^{0}}.
\label{(A15)}
\end{eqnarray}
Explicitly, we find
\begin{eqnarray}
X_{1}&=& \left({z_0 \over 2z_1}{(\gamma-2)\over \gamma}
+{z_{1}\over 2z_0}{(\gamma+2)\over \gamma}\right)
\left(-{\partial \over \partial u}+{\partial \over \partial r}\right)
\nonumber \\
&+& {1 \over 2r}\left[{(z_0)^{2}\over z_1}\left(-{1 \over 2}
-{2 \over \gamma (\gamma+2)}\right)
+z_{1}\left({(\gamma-2)\over 2\gamma}+{1 \over (\gamma-2)}\right)\right]
{\partial \over \partial z_0}
\nonumber \\
&+& {1 \over 2r}\left[{(z_1)^{2}\over z_0}\left(-{1 \over 2}
+{2 \over \gamma (\gamma-2)}\right)
+z_{0}\left({(\gamma+2)\over 2\gamma}-{1 \over (\gamma+2)}\right)\right]
{\partial \over \partial z_1},
\label{(A16)}
\end{eqnarray}
\begin{eqnarray}
X_{2}&=& i \left({z_0 \over 2z_1}{(\gamma-2)\over \gamma}
-{z_{1}\over 2z_0}{(\gamma+2)\over \gamma}\right)
\left({\partial \over \partial u}-{\partial \over \partial r}\right)
\nonumber \\
&+& {i \over 2r}\left[{(z_0)^{2}\over z_1}\left({1 \over 2}
-{2 \over \gamma (\gamma+2)}\right)
+z_{1}\left({(\gamma-2)\over 2\gamma}+{1 \over (\gamma-2)}\right)\right]
{\partial \over \partial z_0}
\nonumber \\
&-& {i \over 2r}\left[{(z_1)^{2}\over z_0}\left({1 \over 2}
-{2 \over \gamma (\gamma-2)}\right)
+z_{0}\left({(\gamma+2)\over 2\gamma}-{1 \over (\gamma+2)}\right)\right]
{\partial \over \partial z_1},
\label{(A17)}
\end{eqnarray}
\begin{equation}
X_{3}={2 \over \gamma}\left(-{\partial \over \partial u}
+{\partial \over \partial r}\right)
+{1 \over 2r}\left(z_0 {\partial \over \partial z_{0}}
-z_1 {(\gamma+2)\over \gamma}{\partial \over \partial z_1}\right).
\label{(A18)}
\end{equation}

The boost ($K_i$) and rotation ($J_{ij}$)  
vector fields for Lorentz transformations
in Minkowski spacetime can be written in $u,r,\xi,{\bar \xi}$ 
coordinates as is shown, for example, in Ref. 
\cite{QG1}. At that stage, by using again Eqs. (4.7), (4.8)
and (A7)-(A10) we find
\begin{eqnarray}
K_{1}&=& {1 \over 2}\left({z_0 \over z_1}{(\gamma-2)\over \gamma}
+{z_{1}\over z_{0}}{(\gamma+2)\over \gamma}\right)
\left(-u {\partial \over \partial u}+(u+r)
{\partial \over \partial r}\right)
\nonumber \\
&-& {(u+r)\over 2r}\left[{(z_0)^{2}\over z_1}\left({1 \over 2}
-{2 \over \gamma(\gamma +2)}\right)-z_{1}\left(
{(\gamma-2)\over 2\gamma}+{1 \over (\gamma-2)}\right)\right]
{\partial \over \partial z_0}
\nonumber \\
&+& \left[{(z_1)^{2}\over z_0} \left({1 \over 2}
-{2 \over \gamma(\gamma-2)}\right)
+z_0 \left(-{(\gamma+2)\over 2\gamma}+{1 \over (\gamma+2)}
\right)\right]{\partial \over \partial z_{1}},
\label{(A19)}
\end{eqnarray}
\begin{eqnarray}
K_{2}&=& {i \over 2}\left({z_0 \over z_1}{(\gamma-2)\over \gamma}
-{z_{1}\over z_{0}}{(\gamma+2)\over \gamma}\right)
\left(u {\partial \over \partial u}-(u+r)
{\partial \over \partial r}\right)
\nonumber \\
&+& i {(u+r)\over 2r} \biggr \{
\left[{(z_0)^{2}\over z_1}\left({1 \over 2}
-{2 \over \gamma(\gamma +2)}\right)+z_{1}\left(
{(\gamma-2)\over 2\gamma}+{1 \over (\gamma-2)}\right)\right]
{\partial \over \partial z_0}
\nonumber \\
&+& \left[{(z_1)^{2}\over z_0} \left(-{1 \over 2}
+{2 \over \gamma(\gamma-2)}\right)
+z_0 \left(-{(\gamma+2)\over 2\gamma}+{1 \over (\gamma+2)}
\right)\right]{\partial \over \partial z_{1}}
\biggr \},
\label{(A20)}
\end{eqnarray}
\begin{equation}
K_3={2 \over \gamma}\left(-u {\partial \over \partial u}
+(u+r){\partial \over \partial r}\right)
+{1 \over 2r}(u+r)\left(z_{0} {\partial \over \partial z_0}
-z_1 {(\gamma+2)\over \gamma}{\partial \over \partial z_1}\right),
\label{(A21)}
\end{equation}
\begin{equation}
J_{12}={i \over 2}\left(z_{0}{\partial \over \partial z_{0}}
-z_{1}{\partial \over \partial z_{1}}\right),
\label{(A22)}
\end{equation}
\begin{eqnarray}
J_{23}&=& {i \over 2}\left[\left(-{(z_0)^{2}\over 2z_{1}}
{\gamma \over (\gamma+2)}+z_{1}\left({1 \over 2}
-{1 \over (\gamma-2)}\right)\right)\right]{\partial \over \partial z_0}
\nonumber \\
&+& {i \over 2}\left[\left(-{(z_1)^{2}\over 2z_0}
{\gamma \over (\gamma-2)}+z_0 \left({1 \over 2}
+{1 \over (\gamma+2)}\right)\right)\right]
{\partial \over \partial z_1},
\label{(A23)}
\end{eqnarray}
\begin{eqnarray}
J_{31}&=& -{1 \over 2}\left[\left({(z_0)^{2}\over 2z_{1}}
{\gamma \over (\gamma+2)}+z_{1}\left({1 \over 2}
-{1 \over (\gamma-2)}\right)\right)\right]{\partial \over \partial z_0}
\nonumber \\
&+& {1 \over 2}\left[\left({(z_1)^{2}\over 2z_0}
{\gamma \over (\gamma-2)}+z_0 \left({1 \over 2}
+{1 \over (\gamma+2)}\right)\right)\right]
{\partial \over \partial z_1}.
\label{(A24)}
\end{eqnarray}
\vskip 100cm

\renewcommand{\theequation}{B.\arabic{equation}}

\leftline {\bf Appendix B: Lie brackets of asymptotic Killing fields}
\setcounter{equation}{0}

\vskip 1cm

Given the vector fields (4.12) and (4.13), the evaluation of their
Lie bracket shows that
\begin{equation}
[\xi_{1},\xi_{2}]=\rho_{1}\left({\partial \over \partial u}
-{\partial \over \partial r}\right)
+\rho_{2}{\partial \over \partial z_{0}}
+\rho_{3}{\partial \over \partial z_{1}},
\label{(B1)}
\end{equation}
where, upon defining the functions
\begin{equation}
\alpha_{1}={2(z_0)^{3}z_1 \over r} \gamma
\left({1 \over 4}-{1 \over \gamma(\gamma+2)}\right),
\label{(B2)}
\end{equation}
\begin{equation}
\alpha_{2}={2 \over \gamma}{1 \over r^{2}}
{(z_0)^{2}\over z_1}\left({1 \over 4}
-{1 \over \gamma(\gamma+2)}\right),
\label{(B3)}
\end{equation}
\begin{equation}
\alpha_{3}={(z_0)^{3}z_{1}\over r} \gamma
\left[{1 \over (\gamma+2)}-{(\gamma+2)\over 2 \gamma}\right],
\label{(B4)}
\end{equation}
\begin{equation}
\alpha_{4}={z_0 \over \gamma}{1 \over r^{2}}
\left[{1 \over (\gamma+2)}-{(\gamma+2)\over 2 \gamma}\right],
\label{(B5)}
\end{equation}
\begin{equation}
\alpha_{5}={z_{0}\over 2r}\left({2 \over \gamma}-1 \right)
\left[{1 \over 2z_1} \left(1-{2 \over \gamma}\right)
+(z_0)^{2}z_{1}\gamma \right],
\label{(B6)}
\end{equation}
\begin{equation}
\alpha_{6}={(z_0)^{2}\over 4 z_1 r^{2}}
\left({2 \over \gamma}-1 \right)^{2},
\label{(B7)}
\end{equation}
\begin{equation}
\alpha_{7}={(z_0)^{4}z_{1}(8+(\gamma-2)\gamma^{2}) \over
4r^{2}(\gamma+2)^{2}},
\label{(B8)}
\end{equation}
\begin{equation}
\alpha_{8}={(z_0)^{4}z_{1}\gamma \over 2r^{2}}
\left[{1 \over (\gamma+2)}-{(\gamma+2)\over 2\gamma}\right],
\label{(B9)}
\end{equation}
\begin{equation}
\alpha_{9}={z_0 \over 4r^{2}}
\left({2 \over \gamma}-1 \right)\left[{1 \over (\gamma+2)}
-{(\gamma+2)\over 2 \gamma}
+{4(z_0)^{2} (z_1)^{2} \gamma(\gamma+1)\over
(\gamma+2)^{2}}\right],
\label{(B10)}
\end{equation}
\begin{equation}
\alpha_{10}={z_0 z_1 \over 4r}
\left({2 \over \gamma}+1 \right)\left[-{1 \over (z_1)^{2}}
\left(1-{2 \over \gamma}\right)
+2(z_0)^{2}\gamma \right],
\label{(B11)}
\end{equation}
\begin{equation}
\alpha_{11}={z_{0}\over 4r^{2}}\left({4 \over \gamma^{2}}-1 \right),
\label{(B12)}
\end{equation}
\begin{equation}
\alpha_{12}={(z_0)^{2}z_{1}\over 2r^{2}}
\left({2 \over \gamma}+1 \right)\left[
-{1 \over (z_1)^{2}} \left({1 \over 4}-
{1 \over \gamma(\gamma+2)}\right)
+{2(z_0)^{2}\gamma (\gamma+1)\over (\gamma+2)^{2}}\right],
\label{(B13)}
\end{equation}
\begin{equation}
\alpha_{13}={(z_0)^{3} (z_1)^{2}\gamma \over r^{2}}
\left({1 \over 4}-{1 \over \gamma(\gamma +2)}\right),
\label{(B14)}
\end{equation}
\begin{equation}
\alpha_{14}={z_{0}(8-(\gamma-6)\gamma^{2})\over
16 r^{2}\gamma^{2}},
\label{(B15)}
\end{equation}
we find that
\begin{equation}
\rho_{1}=\alpha_{1}+\alpha_{3}+\alpha_{5}+\alpha_{10}=0,
\label{(B16)}
\end{equation}
\begin{equation}
\rho_{2}=\alpha_{2}+\alpha_{6}+\alpha_{7}+\alpha_{8}+\alpha_{12}=0,
\label{(B17)}
\end{equation}
\begin{equation}
\rho_{3}=\alpha_{4}+\alpha_{9}+\alpha_{11}+\alpha_{13}+\alpha_{14}=0.
\label{(B18)}
\end{equation}
In the course of performing the calculation, the definition (2.7) leads
to the useful identity
\begin{equation}
{1 \over \gamma^{2}}={4(z_0)^{2} (z_1)^{2}\over (\gamma^{2}-4)}.
\label{(B19)}
\end{equation}

An analogous procedure shows that
\begin{equation}
[\xi_{2},\xi_{3}]=[\xi_{3},\xi_{1}]=0,
\label{(B20)}
\end{equation}
with the help of two additional sets of $14$ nonvanishing functions,
one set for each Lie bracket in (B20). For example, in the Lie bracket
among $\xi_2$ and $\xi_3$, the coefficient of
${\partial \over \partial z_{0}}$ is the function
\begin{eqnarray}
\rho &=& -{z_{0}\over 4r^{2}}
-{(z_0)^{2}\over 16r^{2}}\left(1-{4 \over \gamma(\gamma+2)}\right)
2 z_{0} (z_1)^{2}\gamma \left(1-{\gamma^{2}\over (\gamma-2)^{2}}\right)
\nonumber \\
&+& {1 \over 16 r^{2}}\left(1+{4 \over \gamma(\gamma-2)}\right)
\left[2 z_{0} \left(1-{4 \over \gamma(\gamma+2)}\right)
-2 (z_0)^{3}(z_1)^{2}\gamma \left({\gamma^{2}\over
(\gamma+2)^{2}}-1 \right)\right]
\nonumber \\
&+& {z_0 \over 8r^{2}}\left(1-{4 \over \gamma (\gamma-2)}\right)
\left[\left(1-{4 \over \gamma(\gamma+2)}\right)
+(z_0)^{2} (z_1)^{2} \gamma 
\left({\gamma^{2}\over (\gamma+2)^{2}}-1 \right)\right]
\nonumber \\
&+& {z_0 \over 8r^{2}} 
\left(1+{4 \over \gamma (\gamma+2)}\right)
\left[{1 \over 2}\left(1+{4 \over \gamma (\gamma-2)}\right)
+(z_0)^{2} (z_1)^{2}\gamma 
\left(1-{\gamma^{2}\over (\gamma-2)^{2}}\right)\right]
\nonumber \\
&=& 0.
\label{(B21)}
\end{eqnarray}

\end{document}